\documentclass{article}
\usepackage{color}
\usepackage{graphicx}
\usepackage{amsmath, amsthm}
\usepackage{mathtools}
\usepackage{natbib}
\usepackage{url}
\RequirePackage[colorlinks,citecolor=blue,urlcolor=blue]{hyperref}
\usepackage{xcolor}
\usepackage{chngcntr}
\usepackage{subfig}
\usepackage{hyperref}
\usepackage{algorithm}
\usepackage[noend]{algpseudocode}
\usepackage{setspace}
\newtheorem{theorem}{Theorem}

\usepackage{xfrac}
\definecolor{forest}{rgb}{0.133,0.545,0.133}
\usepackage{multirow}
\usepackage{amsfonts}

\usepackage{enumitem}
\usepackage{caption}
\usepackage{array}
\usepackage{makecell}

\evensidemargin=\oddsidemargin
\usepackage{etoolbox}

\newif\ifabbreviation
\pretocmd{\thebibliography}{\abbreviationfalse}{}{}
\AtBeginDocument{\abbreviationtrue}

\begin{document}
	\newcommand{\bb}{\boldsymbol{\beta}}

	\title{An Efficient Framework for Robust Sample Size Determination}


	\author{Luke Hagar\footnote{Luke Hagar is the corresponding author and may be contacted at \url{l.hagar@uq.edu.au}.} \hspace{35pt} Andrew J. Martin \bigskip \\ 
 \textit{Clinical Trials Capability, The University of Queensland}}

	\date{}

	\maketitle

	\begin{abstract}

In many settings, robust data analysis involves computational methods for uncertainty quantification and statistical inference. To design frequentist studies that leverage robust analysis methods, suitable sample sizes to achieve desired power are often found by estimating sampling distributions of $p$-values via intensive simulation. Moreover, most sample size recommendations rely heavily on assumptions about a single data-generating process. Consequently, robustness in data analysis does not by itself imply robustness in study design, as examining sample size sensitivity to data-generating assumptions typically requires further simulations. We propose an economical alternative for determining sample sizes that are robust to multiple data-generating mechanisms. Applying our theoretical results that model $p$-values as a function of the sample size, we assess power across the sample size space using simulations conducted at only two sample sizes for each data-generating mechanism. We demonstrate the broad applicability of our methodology to study design based on M-estimators in both experimental and observational settings through a varied set of clinical examples.

		\bigskip

	\end{abstract}

	\maketitle

	\baselineskip=19.5pt



 \section{Introduction}\label{sec:intro}

The scientific value of a hypothesis test depends on the selection of a sample size that provides adequate power \citep{cohen1962statistical, lehmann2005testing, faridani2025testing}. In frequentist settings,  hypothesis tests typically rely on $p$-values to  reject a null hypothesis $H_0$. For simple tests, sample sizes with adequate power to reject $H_0$ can be calculated analytically using anticipated standard errors. However, many modern tests rely on computational methods for uncertainty quantification \citep{huber1973robust,efron1982jackknife, laan2003unified, chatterjee2005generalized}. These methods can increase robustness to underlying assumptions \citep{liang1986longitudinal, robins1994estimation} or accommodate complex data structures \citep{laird1982random, stroup2024generalized}. The absence of tractable, closed-form expressions for standard errors renders analytical sample size determination (SSD) impractical for many hypothesis tests. Below, we outline three representative classes of methods -- doubly robust estimators, estimating equations, and generalized linear mixed models -- to illustrate how reliance on computational uncertainty quantification complicates SSD and motivates new design methodology.

Doubly robust estimators \citep{robins1994estimation, bang2005doubly} yield unbiased estimates of causal effects in observational settings when at least one of the following two models is correctly specified: (i) the outcome model or (ii) the propensity score model for the treatment assignment mechanism \citep{rosenbaum1983central, imai2014covariate}. Analytical derivation of standard errors requires knowledge of the true outcome and propensity score models \citep{robins1994estimation}; such derivations are impractical since doubly robust estimators are useful when there is uncertainty regarding one of the two models. Computational strategies (e.g., the bootstrap) facilitate uncertainty quantification in doubly robust analysis \citep{efron1982jackknife}, necessitating simulation-based design. Moreover, sample size recommendations depend strongly on assumptions about both models and the anticipated distribution of covariates. As a result, robustness in data \emph{analysis} does not extend to \emph{design} in the absence of efficient methods for evaluating the impact of data-generating assumptions on SSD.


In the estimating equation framework \citep{godambe1960optimum, liang1986longitudinal}, inference is based on estimating equations that require weaker distributional assumptions than fully parametric likelihoods, enhancing robustness to model misspecification \citep{fitzmaurice2012applied}. Estimating equations are often used to accommodate heterogeneity \citep{parzen1994resampling, pan2001akaike} or correlated responses \citep{zeger1986longitudinal, ito2023grouped} with limited assumptions. To analytically derive sandwich standard errors for data analysis \citep{huber1973robust}, the full data-generation process must be specified \citep{white1980heteroskedasticity}. This requirement precludes analytical SSD since estimating equations are most useful precisely when a full parametric model cannot be confidently specified. Standard errors are therefore estimated computationally using empirical model residuals. Ideally, one would robustly assess how data-generating assumptions -- including those not used for estimation in a semiparametric framework -- impact SSD using efficient simulation.

Generalized linear mixed models (GLMMs) provide a parametric approach to account for heterogeneity and dependence in clustered settings \citep{breslow1993approximate, rue2009approximate}. While these methods require the complete specification of a likelihood function, they are more robust to missing data than estimating equations \citep{little2019statistical}. To derive standard errors at the design stage, one must account for missingness mechanisms and integrate over the distributions of covariates and random effects. These integrals are often intractable and are therefore approximated using Monte Carlo simulation. Existing software -- for instance, the \texttt{SIMR} package in R \citep{green2016simr} -- can be used to conduct power analysis for GLMMs, but these methods rely on naive simulation and inefficiently ignore information from large-sample theory \citep{vaart1998bvm}. 

Thus, it is important to assess how data-generating assumptions impact SSD, where power for a given sample size is determined by estimating the sampling distribution of $p$-values using simulation. In regulatory contexts, the U.S. Food and Drug Administration (FDA) recommends using at least $10^4$ simulation repetitions to estimate sampling distributions for each data-generating scenario considered \citep{fda2019adaptive}. There is a need for an economical framework to robustly select sample sizes that yield sufficient power under a range of plausible data-generating scenarios. 
In this work, we propose a method to assess power across the sample size space by estimating the sampling distribution of $p$-values at only two sample sizes. Related, non-robust methods have been proposed in Bayesian settings for efficient operating characteristic assessment based on sampling distributions of posterior probabilities \citep{hagar2024scalable, hagar2025fdr}; our methodology is fundamentally different because we must develop new theory for frequentist settings, semiparametric models, and point null hypotheses. Our methods are theoretically intricate but simple to implement, promoting an economical and robust framework for SSD that is broadly applicable for use in clustered and non-clustered contexts with hypothesis and equivalence tests \citep{wellek2010testing}.

The remainder of this article is structured as follows. In Section \ref{sec:methods}, we introduce preliminary concepts required to describe our methods. In Section \ref{sec:proxy}, we construct a proxy to the sampling distribution of $p$-values and prove novel theoretical results about this proxy. In Section \ref{sec:power}, we adapt those theoretical results to develop a procedure for robust SSD that requires estimation of the sampling distribution of $p$-values at only two sample sizes for each data-generating model considered. Section \ref{sec:studies} illustrates the use of our method and its performance in the various statistical contexts overviewed above. We conclude with a summary and discussion of extensions to this work in Section \ref{sec:disc}.


\section{Preliminaries}\label{sec:methods} 

Our SSD framework represents data from a random future sample as $\boldsymbol{W}^{_{(n)}}$, where the sample size $n$ is the number of independent sampling units (ISUs). The observed data are denoted by $\boldsymbol{w}^{_{(n)}}$. For non-clustered data, each observation is an ISU. Multiple dependent observations may comprise a single ISU in settings with clustered data. For each ISU, the sample $\boldsymbol{W}^{_{(n)}}$ may consist of response(s) $\boldsymbol{Y}$ along with a vector or matrix of additional covariates $\boldsymbol{X}$ that may encode information about assignment to treatment groups or ISUs. We assume that each ISU in $\boldsymbol{W}^{_{(n)}}$ is generated independently according to some data generation process $\Psi$. This data generation process is specified by choosing parameters $\boldsymbol{\eta}$ for a (semi)parametric model, and additional parameters $\boldsymbol{\rho}$ that are not incorporated into data analysis may also be required to generate covariates, implement censoring, or select ISU sizes. We note that even when using a semiparametric model to analyze data, the data generation process $\Psi$ must be fully specified at the design stage. 

We let the estimand $\theta$ be a function $g(\cdot)$ of the model parameters $\boldsymbol{\eta}$: $\theta = g(\boldsymbol{\eta})$. Upon collecting data, we aim to reject $H_0$ in favour of a complementary alternative hypothesis $H_1$. Our framework accommodates the following types of null hypotheses. First, we consider one-sided hypotheses formulated as $H_0: \theta \le \theta_0$ or $H_0: \theta \ge \theta_0$ for a specified null value $\theta_0$. The $p$-values used to assess these hypotheses are respectively $p(\boldsymbol{w}^{_{(n)}}) = \Pr(T \ge t(\boldsymbol{w}^{_{(n)}})~|~H_0)$ and $p(\boldsymbol{w}^{_{(n)}}) = \Pr(T \le t(\boldsymbol{w}^{_{(n)}})~|~H_0)$, where $t(\boldsymbol{w}^{_{(n)}})$ is an observed test statistic and $T$ denotes its null sampling distribution. In this work, we restrict attention to test statistics whose distributions are derived from the sampling distribution of an M-estimator \citep{huber1964robust}. Because maximum likelihood estimators are M-estimators \citep{vaart1998bvm}, this consideration encompasses fully parametric inference via maximum likelihood estimation.  Second, we consider equivalence tests such that $H_0: \theta \le \theta_{0,L} \cup \theta \ge \theta_{0,U}$, where $\theta_{0,L}$ and $\theta_{0,U}$ are lower and upper endpoints for an interval of practical equivalence. The $p$-value for such hypotheses is the larger of the two one-sided $p$-values corresponding to $H_0: \theta \le \theta_{0,L}$ and $H_0: \theta \ge \theta_{0,U}$ \citep{schuirmann1987comparison}. Third, we accommodate two-sided hypotheses such that $H_0 = \theta_0$. For an appropriately centered test statistic, the $p$-value is $p(\boldsymbol{w}^{_{(n)}}) = 2\Pr(T \ge \lvert t(\boldsymbol{w}^{_{(n)}}) \rvert~|~H_0)$. Henceforth, we generally denote the $p$-value as $p(\boldsymbol{w}^{_{(n)}})$ with its explicit form determined by the specification of $H_0$.

Algorithm \ref{alg.int} details a straightforward procedure to estimate the sampling distribution of $p$-values using simulation at a given sample size $n$. To conduct SSD, we specify a data-generating process $\Psi_1$ such that $H_1$ is true. We later examine a collection of data-generating processes $\{\Psi_{1,k}\}_{k=1}^K$ for robust SSD, but we begin by considering a single process $\Psi_1$. We generate samples from $\Psi_1$ across simulation repetitions $r = 1, \dots, R$ in Line 3. For each generated sample, we compute the $p$-value in Line 4. This step may involve estimating standard errors using computational methods as discussed in Section \ref{sec:intro}. 

\begin{algorithm}
\caption{Sampling Distribution Estimation}
\label{alg.int}

\begin{algorithmic}[1]
\setstretch{1}
\Procedure{Estimate}{$H_0$, $\Psi_1$, $g(\cdot)$, $n$, $R$}
\For{$r$ in 1:$R$}
    \State Generate $\boldsymbol{w}^{_{(n)}}_{r} \sim \Psi_1$ 
    \State Compute estimate $\hat{p}(\boldsymbol{w}^{_{(n)}}_{r})$ 
    \EndFor
    \State \Return $\{\hat{p}(\boldsymbol{w}^{_{(n)}}_{r})\}_{r = 1}^R$
\EndProcedure

\end{algorithmic}
\end{algorithm}

The collection of estimates $\{\hat{p}(\boldsymbol{w}^{_{(n)}}_{r})\}_{r = 1}^R$ returned by Algorithm \ref{alg.int} estimates the sampling distribution of $p$-values under $\Psi_1$. Given a desired type I error rate $\alpha$, power under $\Psi_1$ is estimated as
\begin{equation}\label{eq:oc.est}
\dfrac{1}{R}\sum_{r=1}^R\mathbb{I}\left\{\hat{p}(\boldsymbol{w}^{_{(n)}}_{r}) \le \alpha\right\}.
\end{equation} 
For a given choice of $\Psi_1$, power increases with $n$. To determine whether the target power for the study is attained under $\Psi_1$, we introduce the notation $\xi(a,b)$ to denote the $a^{\text{th}}$ smallest order statistic of the collection of observations $b$. For a given sample size $n$, the estimated power under $\Psi_1$ is at least $1 - \beta$ if and only if $\xi_1 = \xi(\lceil  (1-\beta) R \rceil, \{\hat{p}(\boldsymbol{w}^{_{(n)}}_{r})\}_{r = 1}^R) \le \alpha$. 


Our focus is proposing an SSD framework that robustly ensures sufficient power across various data-generation processes $\{\Psi_{1,k}\}_{k=1}^K$. This framework identifies the smallest sample size $n$ such that for 
all $\Psi_{1,k}$, the corresponding quantities $\{\xi_{1,k}\}_{k=1}^K$ do not exceed $\alpha$. Although multiple data-generating processes are considered at the design stage, no multiple-testing adjustment is required since the observed data will ultimately be analyzed using a single hypothesis test. In principle, robust selection of the sample size $n$ requires independently implementing Algorithm \ref{alg.int} for each combination of $n$ and $\Psi_{1, k}$ under consideration to estimate the corresponding sampling distributions of $p$-values. This approach is computationally intensive. However, the sample size space can be explored more efficiently by leveraging estimated sampling distributions from previously considered sample sizes under a particular $\Psi_{1,k}$ to approximate  $\xi_{1,k}$ at new values of $n$. We could use this process to explore each $\Psi_{1,k}$ with substantially fewer simulation repetitions. We propose such a method for robust SSD in this paper and begin its formal development in Section \ref{sec:proxy}. 

    \section{A Proxy for the Sampling Distribution of \texorpdfstring{$p$}{p}-Values}\label{sec:proxy} 

     To determine sample sizes for complex hypothesis tests, we must estimate the sampling distribution of $p$-values across a range of sample sizes $n$. We approximate these sampling distributions by generating data $\boldsymbol{w}^{_{(n)}}$ using Algorithm \ref{alg.int}. For theoretical development, we introduce a proxy for these sampling distributions. Although  this proxy is not directly used in the design methods proposed in Section \ref{sec:power}, it provides theoretical justification for our methodological development. A thorough understanding of this proxy is not required to apply our design methods that are simple to implement.

     Our proxy construction relies on the regularity conditions in Appendix A of the supplement. Appendix A summarizes the conditions for the asymptotic normality of the M-estimator \citep{vaart1998bvm}. To construct the proxy distribution, we consider the sampling distribution of the M-estimator under a single data-generation process $\Psi_1$ such that $H_1$ is true. This process $\Psi_1$ is characterized by model parameters $\boldsymbol{\eta}_1$, additional parameters $\boldsymbol{\rho}_1$, and the associated estimand $\theta_1 = g(\boldsymbol{\eta}_1)$. When the conditions in Appendix A are satisfied, the approximate sampling distribution of the M-estimator $\hat{\theta}^{_{(n)}}~|~\boldsymbol{\eta} = \boldsymbol{\eta}_1$ is $\mathcal{N}(\theta_{1}, n^{-1}\lambda^2(\boldsymbol{\eta}_1, \boldsymbol{\rho}_1))$, where $n$ is the number of ISUs. The quantity $ \lambda^2(\boldsymbol{\eta}_1, \boldsymbol{\rho}_1) = \lambda^2_1$ depends on the underlying (semi)parametric model and additional  parameters; concrete examples are provided in Section \ref{sec:studies}. For theoretical purposes only, a realization from this normal sampling distribution could be generated using cumulative distribution function (CDF) inversion and a point $u_r \in [0,1]$:
   \begin{equation}\label{eq:cdf.inv}
\hat{\theta}^{_{(n)}}_{r} = \theta_{1} + \Phi^{-1}(u_{r})\dfrac{\lambda_1}{\sqrt{n}},
\end{equation} 
where $\Phi(\cdot)$ is the standard normal CDF.
          
Implementing this procedure with a sequence of $R$ points $\{u_{r}\}_{r = 1}^R \sim \mathcal{U}([0,1])$ yields a simulated sample from the approximate sampling distribution of $\hat{\theta}^{_{(n)}}$ under $\Psi_1$. We now use this simulated sample to construct a large-sample proxy to the sampling distribution of $p$-values. We begin by considering the $p$-value for the one-sided hypothesis $H_0: \theta \le \theta_0$. We derive a proxy $p$-value corresponding to the simulated value $\hat{\theta}^{_{(n)}}_{r}$ as
\begin{equation}\label{eq:proxy.1}
\begin{aligned}
p^{_{(n)}}_r &= \Pr(\hat{\Theta}^{_{(n)}} \ge \hat{\theta}^{_{(n)}}_{r}~|~H_0) \\
     &= \Pr\left(\theta_0 + Z\dfrac{\lambda_0}{\sqrt{n}} \ge \theta_1 + \Phi^{-1}(u_r)\dfrac{\lambda_1}{\sqrt{n}}\right) \\
      &= 1 - \Phi\left(a_1 \sqrt{n} +b_1(u_r)\right),
\end{aligned}
\end{equation}
where $\hat{\Theta}^{_{(n)}}$ is a random variable for $\hat{\theta}^{_{(n)}}$ and $Z$ is a standard normal random variable. The second equality in (\ref{eq:proxy.1}) holds true because under the conditions in Appendix A, the sampling distribution of $\hat{\theta}^{_{(n)}}~|~\boldsymbol{\eta} = \boldsymbol{\eta}_0$ is $\mathcal{N}(\theta_{0}, n^{-1}\lambda^2_0)$, where $\lambda^2_0$ is based on a data-generation process $\Psi_0$ such that $\theta = \theta_0$. The final line in (\ref{eq:proxy.1}) follows from algebra, where $a_1 = (\theta_1 - \theta_0)/\lambda_0$ and $b_1(u_r) = \Phi^{-1}(u_r)\lambda_1/\lambda_0$. 

For the one-sided hypothesis $H_0: \theta \ge \theta_0$, we can follow a similar process to derive a proxy $p$-value corresponding to $\hat{\theta}^{_{(n)}}_{r}$ as
\begin{equation}\label{eq:proxy.2}
\begin{aligned}
p^{_{(n)}}_r &= \Pr(\hat{\Theta}^{_{(n)}} \le \hat{\theta}^{_{(n)}}_{r}~|~H_0) \\
      &= \Phi\left(a_1 \sqrt{n} +b_1(u_r)\right).
\end{aligned}
\end{equation}
For equivalence tests with $H_0: \theta \le \theta_{0,L} \cup \theta \ge \theta_{0,U}$, the proxy $p$-value corresponding to $\hat{\theta}^{_{(n)}}_{r}$ is the maximum of the proxy $p$-value in (\ref{eq:proxy.1}) when $\theta_0 = \theta_{L,0}$ and the proxy $p$-value in (\ref{eq:proxy.2}) when $\theta_0 = \theta_{U,0}$. Theoretical consideration of proxy $p$-values for equivalence tests can therefore be conducted by examining the proxy $p$-values in (\ref{eq:proxy.1}) and (\ref{eq:proxy.2}). 

Lastly, we consider the two-sided hypothesis $H_0: \theta = \theta_0$. The proxy $p$-value corresponding to $\hat{\theta}^{_{(n)}}_{r}$ is
\begin{equation}\label{eq:proxy.3}
\begin{aligned}
p^{_{(n)}}_r &= \Pr(\hat{\Theta}^{_{(n)}} - \theta_0 \le -\lvert \hat{\theta}_r^{_{(n)}} - \theta_0 \rvert~|~H_0) + \Pr(\hat{\Theta}^{_{(n)}} - \theta_0 \ge \lvert \hat{\theta}^{_{(n)}}_r - \theta_0 \rvert~|~H_0) \\
&=  \Pr\left(Z\dfrac{\lambda_0}{\sqrt{n}} \le -\left\lvert \theta_1 + \Phi^{-1}(u_r)\dfrac{\lambda_1}{\sqrt{n}} - \theta_0 \right\rvert\right) +  \Pr\left(Z\dfrac{\lambda_0}{\sqrt{n}} \ge \left\lvert \theta_1 + \Phi^{-1}(u_r)\dfrac{\lambda_1}{\sqrt{n}} - \theta_0 \right\rvert\right) \\
      &= 2\Phi\left(-\left\lvert a_1 \sqrt{n} +b_1(u_r)\right\rvert\right).
\end{aligned}
\end{equation}
The final equality of (\ref{eq:proxy.3}) can be shown using algebra. While the form of the proxy $p$-value depends on how $H_0$ is specified, we note that the proxy $p$-values in (\ref{eq:proxy.1}), (\ref{eq:proxy.2}), and (\ref{eq:proxy.3}) share a similar structure in which there is a linear function of $\sqrt{n}$ inside a standard normal CDF. The collection of $\{p^{_{(n)}}_{r}\}_{r = 1}^R$ values corresponding to $\{u_{r}\}_{r = 1}^R \sim \mathcal{U}([0,1])$ define our proxy to the sampling distribution of $p$-values under $\Psi_1$. We acknowledge that this proxy to the sampling distribution relies on asymptotic results, so it may differ from the true sampling distribution of $p$-values for finite $n$. 

The proxy sampling distribution therefore only motivates our theoretical result in Theorem \ref{thm1}. This result guarantees that the logit of $p^{_{(n)}}_{r}$ is an approximately linear function of $n$. We later adapt this result to assess the power of a study across a broad range of sample sizes by estimating the true sampling distribution of $p$-values at only \emph{two} values of $n$ for each $\Psi_{1, k}$ process considered. Each $p^{_{(n)}}_{r}$ value in the proxy sampling distribution depends on the value for $\hat{\theta}^{_{(n)}}_{r}$, which depends on the number of ISUs $n$ and the point $u_{r}$. We emphasize that $u_{r} \sim \mathcal{U}([0,1])$ is a stochastic conduit for the data. In Theorem \ref{thm1}, we fix the $u_{r}$ value to explore the behavior of $p^{_{(n)}}_{r}$ as a deterministic function of $n$.

\begin{theorem}\label{thm1}
    Let the conditions in Appendix A be satisfied. Define $\emph{logit}(x) = \emph{log}(x) - \emph{log}(1-x)$. We consider a given point $u_{r} \in [0,1]$ and data-generation process $\Psi_1$. Under these assumptions,
    \vspace*{0pt}
 \begin{enumerate}
     \item[(a)] the functions $p^{_{(n)}}_{r}$ in (\ref{eq:proxy.1}) and (\ref{eq:proxy.2}) are such that $\lim\limits_{n \rightarrow \infty} \dfrac{d}{dn}~\emph{logit}\left(p^{_{(n)}}_{r}\right)= -\dfrac{a_1^2}{2}$. 
     \item[(b)] the function $p^{_{(n)}}_{r}$ in (\ref{eq:proxy.3}) is such that $\lim\limits_{n \rightarrow \infty} \dfrac{d}{dn}~\emph{logit}\left(0.5 p^{_{(n)}}_{r}\right)= -\dfrac{a_1^2}{2}$. 
 \end{enumerate}
\end{theorem} 

We prove parts $(a)$ and $(b)$ of Theorem \ref{thm1} in Appendices B.1 and B.2. We now discuss the practical implications of this theorem. The limiting derivatives in parts $(a)$ and $(b)$ do not depend on $n$. For the proxy $p$-values in (\ref{eq:proxy.1}) and (\ref{eq:proxy.2}), the linear approximation to $l^{_{(n)}}_{r} = \text{logit}(p^{_{(n)}}_{r})$ as a function of $n$ is thus a good global approximation for large values of $n$. This linear approximation should be locally suitable for a range of smaller sample sizes. Therefore, the \emph{quantiles} of the sampling distribution of $l^{_{(n)}}_{r}$ change linearly as a function of $n$ for a given choice of $\Psi_1$. These linear trends also hold true for $\text{logit}(0.5p^{_{(n)}}_{r})$ from (\ref{eq:proxy.3}) with two-sided hypotheses -- i.e., the contribution to the proxy $p$-value from one tail of the sampling distribution. In Section \ref{sec:power}, we exploit and adapt these linear trends in the proxy sampling distribution to flexibly model the logits of $p$-values using linear functions of $n$ when independently simulating samples from a set of $\Psi_1$ processes. While the proxy sampling distribution is predicated on asymptotic results, we illustrate the good performance of our robust SSD procedure with finite sample sizes $n$ in Section \ref{sec:studies}.

   \section{Robust Sample Size Determination Procedure}\label{sec:power}

We generalize the results from Theorem \ref{thm1} to develop an approach in Algorithm \ref{alg2} that is easily implemented and performs well with moderate to large $n$. Algorithm \ref{alg2} helps users efficiently explore the sample size space for each data-generating  process to find the minimum $n$ for which the target power is attained. Our method involves estimating the sampling distributions of $p$-values for each data-generating mechanism at only two sample sizes. The initial sample size $n_0$ can be selected based on the anticipated budget for the study. In Algorithm \ref{alg2}, we add a second subscript $k$ to the samples $\boldsymbol{w}^{_{(n)}}_{r, k}$ and corresponding $p$-values to distinguish between data-generating processes in $\{\Psi_{1, k}\}_{k=1}^K$. Algorithm \ref{alg2} details a general application of our methodology for use with one-sided hypotheses, and we later describe modifications to accommodate equivalence tests and two-sided hypotheses. 

\begin{algorithm}
\caption{Procedure to Robustly Determine Sample Size}
\label{alg2}

\begin{algorithmic}[1]
\setstretch{1}
\Procedure{RobustSSD}{$\{\Psi_{1,k}\}_{k=1}^K$, $g(\cdot)$, $H_0$, $\alpha$, $\beta$, $R$, $n_0$}
\For{$k$ in $1$:$K$}
\State Estimate $\{\widehat{p}(\boldsymbol{w}^{_{(n_0)}}_{r, k})\}_{r = 1}^R$ via Algorithm \ref{alg.int} and their logits $\{\hat{l}^{_{(n_0)}}_{r, k}\}_{r = 1}^R$
\State If $R^{-1}\sum_{r=1}^R\mathbb{I}\{\widehat{p}(\boldsymbol{w}^{_{(n_0)}}_{r, k}) \le \alpha\} \ge 1-\beta$, choose $n_{1,k} < n_0$. If not, choose $n_{1, k} > n_0$.
\State Estimate $\{\widehat{p}(\boldsymbol{w}^{_{(n_{1, k})}}_{r, k})\}_{r = 1}^R$ via Algorithm \ref{alg.int} and their logits $\{\hat{l}^{_{(n_{1,k})}}_{r, k}\}_{r = 1}^R$
\For{$r$ in $1$:$R$}
  \State Join the $r^{\text{th}}$ order statistics of $\{\hat{l}^{_{(n_{0})}}_{s, k}\}_{s = 1}^R$ and $\{\hat{l}^{_{(n_{1,k})}}_{s, k}\}_{s = 1}^R$ with a line to obtain $\hat{l}^{_{(n)}}_{r, k}$ for new $n$ \linebreak \hspace*{42pt}  values.
 \EndFor
 \State Obtain $\{\hat{p}^{_{(n)}}_{r, k}\}_{r = 1}^R$ as the inverse logits of the estimates $\{\hat{l}^{_{(n)}}_{r, k}\}_{r = 1}^R$.
\State Find $n_{2, k}$, the smallest $n \in \mathbb{Z}^+$ such that $R^{-1}\sum_{r=1}^R\mathbb{I}\{\hat{p}^{_{(n)}}_{r, k} \le \alpha\} \ge 1-\beta$.
 \EndFor

 \State \Return maximum of $\{n_{2, k}\}_{k=1}^K$ as recommended $n$

\EndProcedure

\end{algorithmic}
\end{algorithm}

We now overview several of the steps in Algorithm \ref{alg2}. The outer for loop iterates over the data-generating processes and obtains a sample size recommendation for each. In Line 4, we choose a second sample size $n_{1, k}$ at which to estimate the sampling distribution of $p$-values under the process $\Psi_{1, k}$. This sample size could be selected in several ways. If it is possible to estimate the limiting slopes from Theorem \ref{thm1}, logits $\{\hat{l}^{_{(n)}}_{r, k}\}_{r=1}^R$ could initially be estimated for new $n$ values using lines that pass through the points $\{(n_0, \hat{l}^{_{(n_0)}}_{r, k})\}_{r=1}^m$ with these limiting slopes. The value for $n_{1, k}$ can then be found as the smallest sample size such that power based on those estimated logits is at least $1 - \beta$. An alternative strategy is to select $n_0$ and $n_1 = \{n_{1,k}\}_{k=1}^K$ to facilitate visualization of the $\Psi_{1,k}$-specific power curves over a range of relevant sample sizes. 

 We then construct linear approximations separately for each $\Psi_{1,k}$. These approximations use the logits $\{\hat{l}^{_{(n_0)}}_{r, k}\}_{r = 1}^R$ and $\{\hat{l}^{_{(n_{1,k})}}_{r, k}\}_{r = 1}^R$ in Line 7 of Algorithm \ref{alg2}. For one-sided hypotheses, we use these linear approximations to estimate logits of $p$-values for new values of $n$. For equivalence tests, we modify the process in Line 7 to construct linear approximations for the two $p$-values corresponding to $H_0: \theta \le \theta_{0,L}$ and $H_0: \theta \ge \theta_{0,U}$. For two-sided hypotheses, we must divide the $p$-values from Lines 3 and 5 by 2 before creating linear approximations in Line 7.

Given the linear trend in the proxy sampling distribution quantiles discussed in Section \ref{sec:proxy}, it is reasonable to construct these linear approximations based on order statistics of the true sampling distribution estimates. In Line 8 of Algorithm \ref{alg2}, we transform these logits obtained with linear approximations back to the probability scale. For equivalence tests, we modify Line 8 to take the $p$-value as the larger of the probabilities obtained by applying the inverse-logit transformation to the two one-sided logits. For two-sided hypotheses, we multiply the probabilities obtained after applying the inverse-logit transformation in Line 8 by 2. We find the smallest value of $n$ such that the power estimate under $\Psi_{1,k}$ is at least $1-\beta$ in Line 9. In Line 10, we return the maximum sample size recommendation across $\{\Psi_{1,k}\}_{k=1}^K$ to ensure that the sample size is robust to all data-generating mechanisms considered. In Section \ref{sec:studies}, we consider the performance of Algorithm \ref{alg2} in various clinical contexts.

    \section{Numerical Studies}\label{sec:studies}

        \subsection{Doubly Robust Inference}

    We illustrate our method using an example inspired by a target trial emulation that examined whether concomitant docetaxel improves survival among enzalutamide-treated patients with metastatic hormone-sensitive prostate cancer \citep{soon2025target}. The emulation was based on data from the ENZAMET trial \citep{davis2019enzalutamide, sweeney2023testosterone}, which randomized treatment assignment to enzalutamide or standard therapy, with the decision to administer concomitant docetaxel specified prior to randomization.

    In contrast to the retrospective analysis of \cite{soon2025target}, we consider the \emph{a priori} design of a target trial emulation using non-randomized data, where treatment assignment is observational and sample size must be specified in advance. Power assessment in this setting is challenging because inference relies on semiparametric causal estimators with computational uncertainty quantification, making it a natural context for illustrating our robust and economical SSD approach.

    The difference in 5-year overall survival between enzalutamide patients who received docetaxel and those who did not is our estimand $\theta$. We estimate survival differences in a causal inference framework using augmented inverse probability weighting (AIPW) \citep{wang2001augmented}. The propensity score is modeled using logistic regression, while survival and censoring times are modeled using semiparametric Cox proportional hazards models. In this example, estimation of the censoring model is required to compute the inverse probability weights. More details on the AIPW estimator for $\theta$ are provided in Appendix C.1 of the supplement; this M-estimator is doubly robust if (i) the propensity score and censoring time models are correctly specified or (ii) the survival time model is correctly specified. We aim to reject the null hypothesis $H_0: \theta = 0$ in favour of $H_1: \theta \ne 0$. We use a point-null hypothesis for illustration; one-sided and equivalence hypotheses are considered in subsequent examples. We want a type I error rate of $\alpha = 0.1$ and study power of $1 - \beta = 0.8$. 

    While $\theta$ is a function of the survival model parameters $\boldsymbol{\eta}$, additional parameters $\boldsymbol{\rho}$ are required to fully specify the data-generating process, including covariate generation, treatment assignment via the propensity score, and missing-at-random censoring. We consider six data-generating processes $\{\Psi_{1,k}\}_{k=1}^6$ to illustrate common modifications that may arise in causal inference settings. An overview of these processes is given below, with full specification provided in Appendix C.2.  The process $\Psi_{1,1}$ has a Weibull baseline hazard. In $\Psi_{1,2}$ and $\Psi_{1,3}$, $\Psi_{1,1}$ is modified so that the true propensity score and survival time models correspond to simpler models nested within the fitted model. The process $\Psi_{1,4}$ modifies the covariate distribution from $\Psi_{1,1}$, thereby altering treatment assignment induced by the propensity score. Processes $\Psi_{1,5}$ and $\Psi_{1,6}$ modify $\Psi_{1,1}$ so the baseline hazard is lognormal and piecewise constant. All $\{\Psi_{1,k}\}_{k=1}^6$ are such that roughly 10\% of deaths occurring before 5 years are right-censored.

\begin{figure}[!tb]
		\includegraphics[width = \textwidth]{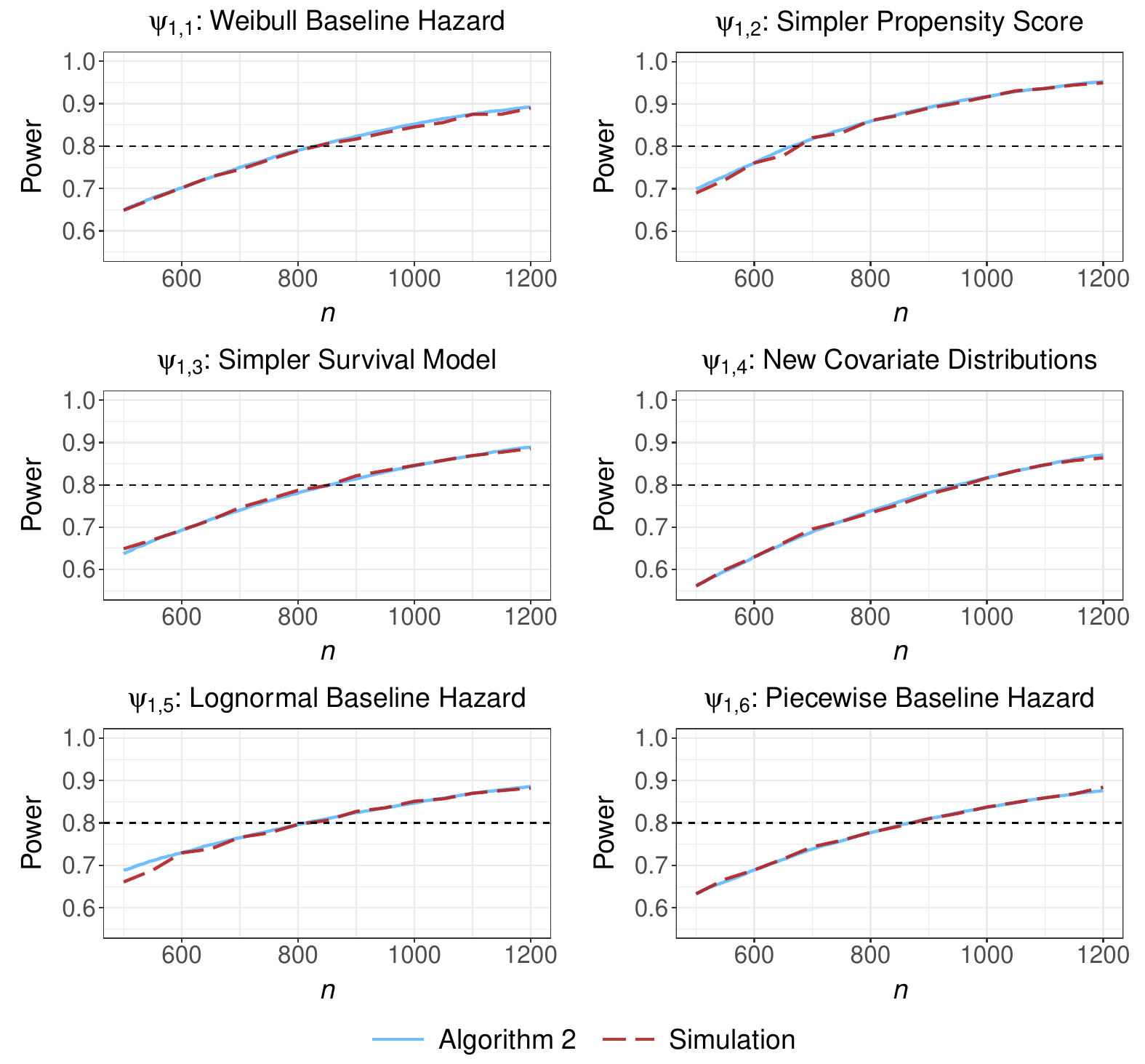} 

		\caption{\label{fig:dr} Power curves obtained using two estimation methods for the doubly-robust survival example. The horizontal dotted line represents $1 - \beta = 0.8$. } 
	\end{figure}

   In all simulations, we estimate the standard error of the AIPW estimator using the nonparametric bootstrap with 100 resamples to reduce the computational overhead. Given the small number of resamples, we approximate the standard error using a function of the median absolute difference \citep{boyles1997estimating}. The uncertainty associated with estimating the inverse probability weights is incorporated into our simulations. Nevertheless, this uncertainty vanishes as $n \rightarrow \infty$ and does not impact our asymptotic results about proxy sampling distributions in Section \ref{sec:proxy}. We first estimate the power curves for each $\Psi_{1,k}$ using Algorithm \ref{alg2} with $n_0 = 600$ and $n_1 = 1100$ corresponding to sample sizes similar to those in the ENZAMET trial. We then approximated the power curve for each $\Psi_{1,k}$ by naively simulating the sampling distribution of $p$-values for $n \in \{500, 550, \dots, 1200\}$. All sampling distributions in this paper were estimated with $R = 10^4$ simulation repetitions. Figure \ref{fig:dr} visualizes the resulting approximated power curves.

    For all $\Psi_{1,k}$ in Figure \ref{fig:dr}, the results from our linear approximations in Algorithm \ref{alg2} align well with those obtained using naive simulation. Even though the naive power curves are subject to simulation variability, we use them as surrogates for the true power. Roughly 30 minutes on a high-computing server with 72 cores were required to approximate the power curve via Algorithm \ref{alg2} for each $\Psi_{1,k}$. Using the same computing resources, naive simulation required approximately 4 hours to approximate the power curves in Figure \ref{fig:dr}. Algorithm \ref{alg2} is more efficient because we need only estimate the sampling distribution of $p$-values at two values of $n$. The robust sample size recommendation obtained via Algorithm \ref{alg2} is $n = 951$ since $\{n_{2,k}\}_{k=1}^6 = \{830, 664, 855, 951, 818, 868 \}$. The processes  $\{\Psi_{1,k}\}_{k=1}^6$ illustrate that modest changes to the data-generation mechanism may result in large changes to the required sample size. In Appendix C.3, we confirm via simulation that the 
    recommended design approximately attains the target type I error rate and power.

    \subsection{Linear Mixed Models}

     We next demonstrate our method by designing a hypothetical health-related quality of life (HRQoL) sub-study inspired by the INTEGRATE I phase II trial \citep{martin2025health}, which randomized patients with advanced gastric or oesophagogastric junction cancer in a 2:1 ratio to receive regorafenib or placebo. In our hypothetical design, the HRQoL sub-study -- and any tolerability issues it uncovers -- would be central to informing go/no-go development decisions. The EORTC Global Health Status/Quality of Life (GHS-QoL) score \citep{fayers2001eortc}, which ranges from 0 to 100 with higher scores denoting better HRQoL, is selected as the primary measure. The longitudinal structure of these data makes linear mixed models a natural choice for analysis. The coefficient for the treatment variable represents the difference in HRQoL for the regorafenib and placebo groups and defines our estimand $\theta$. 

     To analyze this longitudinal outcome, our linear mixed model includes fixed effects for the baseline GHS-QoL score, the treatment assignment, and time since baseline (in weeks). We assume patients are assessed every four weeks after baseline for a maximum of 24 weeks. Our modeling approach accommodates random intercepts and random slopes for the time covariate, with more details provided in Appendix D.1 of the supplement. We aim to reject the null hypothesis $H_0: \theta \le -10$ in favour of $H_1: \theta > -10$. Rejecting $H_0$ allows us to conclude that regorafenib is non-inferior to the placebo with respect to GHS-QoL score when the minimal clinically important difference is 10 points \citep{martin2025health}. For illustrative purposes, we desire a type I error rate of $\alpha = 0.025$ and trial power of $1 - \beta = 0.9$. 

     For this example, we again require additional parameters $\boldsymbol{\rho}$ to generate covariates and implement missing-at-random dropout. We consider six candidate data-generation processes $\{\Psi_{1,k}\}_{k=1}^6$ that are fully specified in Appendix D.2. For all data-generation processes, dropout is monotonic in the sense that participants are not observed at subsequent visits after missing one. In the base-case process $\Psi_{1,1}$, the random-slope variance is set just large enough to avoid convergence issues during model fitting, and the dropout probability at the current visit depends only on the observed GHS-QoL score from the previous visit. In $\Psi_{1,2}$, $\Psi_{1,1}$ is modified so that the random slopes have a larger variance. The processes $\Psi_{1,3}$ and $\Psi_{1,4}$ introduce more complicated dropout mechanisms than $\Psi_{1,1}$, with dropout probabilities additionally depending on time and the treatment assignment to, respectively, induce greater dropout at early visits and in the placebo arm. In $\Psi_{1,5}$, dropout depends only on the baseline GHS-QoL score. The process $\Psi_{1,6}$ modifies $\Psi_{1,1}$ so the random intercepts have a larger variance. All $\{\Psi_{1,k}\}_{k=1}^6$ are such that roughly 75\% of patients dropout before their final visit at 24 weeks.

\begin{figure}[!tb]
		\includegraphics[width = \textwidth]{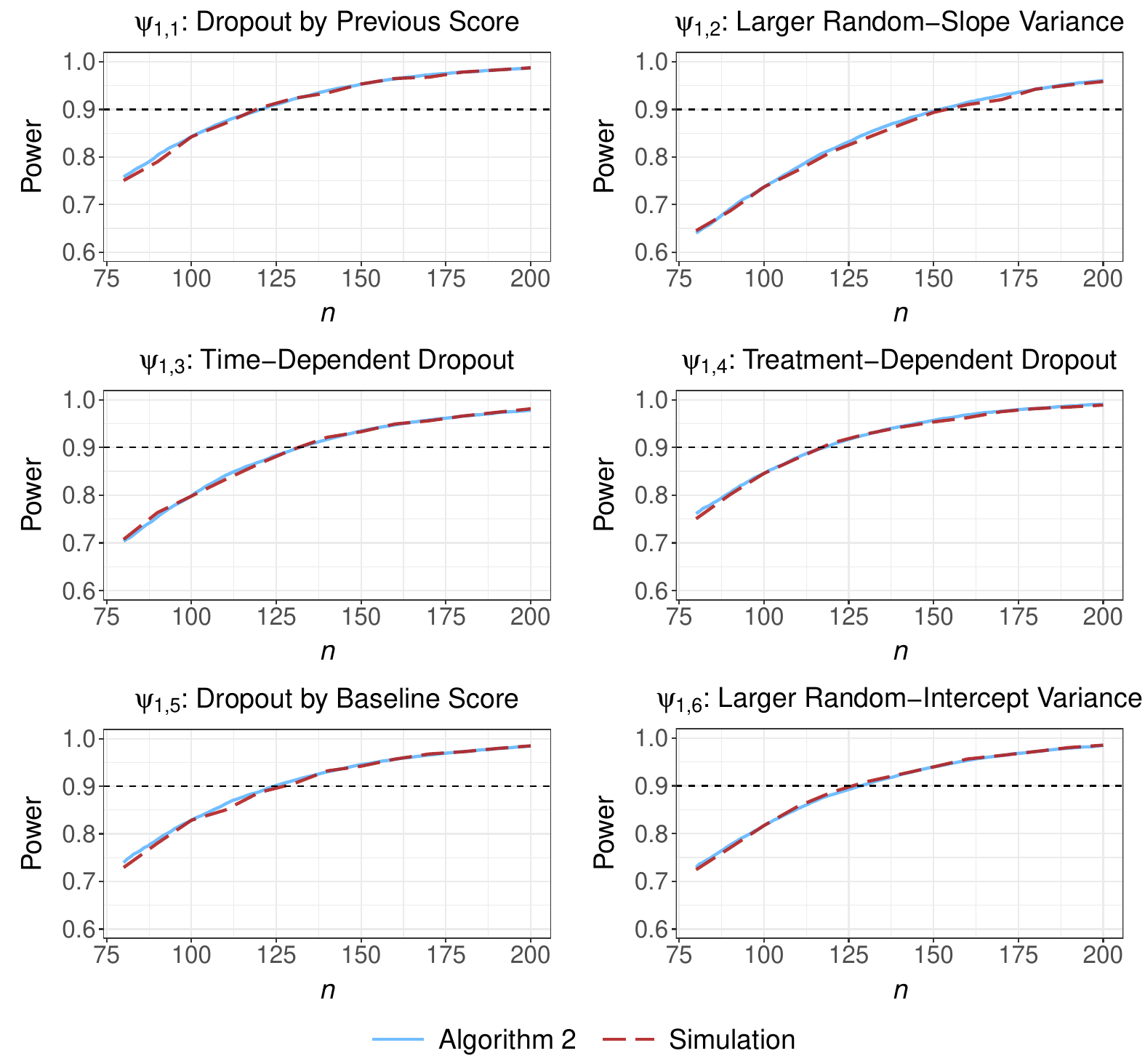} 

		\caption{\label{fig:lmm} Power curves obtained using two estimation methods for the example with linear mixed models. The horizontal dotted line represents $1 - \beta = 0.9$. } 
	\end{figure}

     In all simulations, standard errors for $\hat{\theta}$ are obtained from linear mixed models fit using restricted maximum likelihood. Estimating sampling distributions of $p$-values via simulation allows us to obtain power estimates that account for the covariate distributions and dropout mechanisms. We first estimate the power curves for each $\Psi_{1,k}$ using Algorithm \ref{alg2} with $n_0 = 100$ and $n_1 = 180$ to explore sample sizes similar to that from INTEGRATE 1. We then approximated the power curve for each $\Psi_{1,k}$ by naively simulating the sampling distribution of $p$-values for $n \in \{80, 90, \dots, 200\}$. Figure \ref{fig:lmm} visualizes the power curves.

    For all $\Psi_{1,k}$ in Figure \ref{fig:lmm}, the results from our linear approximations in Algorithm \ref{alg2} again align well with those obtained using naive simulation. Roughly 30 seconds on a high-computing server with 72 cores were required to approximate the power curve via Algorithm \ref{alg2} for each $\Psi_{1,k}$. It took about 4 minutes to approximate the power curves in Figure \ref{fig:lmm} with naive simulation. The robust sample size recommendation obtained via Algorithm \ref{alg2} is $n = 152$ since $\{n_{2,k}\}_{k=1}^6 = \{121, 152, 132, 118, 125, 129 \}$. In Appendix D.3, we confirm via simulation that the desired type I error rate and power are reasonably attained for the recommended design.

    \subsection{Generalized Estimating Equations}

    Lastly, we use our method to design a hypothetical equivalence study comparing an established intravenous anti-seizure drug with a new oral formulation containing the same active ingredient. The data structure and outcome model are informed by the seizure trial described in \citet{thall1990some}. The objective of the study is to determine whether the two formulations provide practically equivalent control of seizure rates under maintenance therapy. Demonstrating equivalence would justify extrapolation of the existing efficacy and safety evidence from the intravenous formulation to the oral formulation, supporting clinical interchangeability.
        
    Prior to baseline, the number of seizures experienced in the previous 8 weeks is recorded for each patient. The seizure counts for all patients are also recorded for four 2-week periods after baseline. We use a generalized estimating equation (GEE) with a Poisson mean model to conduct inference about the marginal seizure rates under the two formulations, accounting for within-patient correlation across repeated measurements.

    As detailed in Appendix E.1 of the supplement, one regression coefficient in the model is the log-ratio of the expected seizure rates in the new formulation group in the post-baseline period, as compared to the reference formulation group, in the same period. This regression coefficient is our estimand $\theta$. In this example, we consider an equivalence test and aim to reject the composite null hypothesis $H_0: \theta \le -\log(4/3) \cup \theta \ge \log(4/3)$ in favour of $H_1: -\log(4/3) < \theta < \log(4/3) $. This interval of practical equivalence is chosen solely for illustration to evaluate the performance of our method in settings with less than $n = 100$ ISUs, rather than to reflect clinical or regulatory judgement. For this example, we desire a type I error rate of $\alpha = 0.05$ and trial power of $1 - \beta = 0.8$. 

    The data-generation processes we consider require additional parameters not included in the GEE model to induce overdispersion and various dependence structures between seizure counts from the same patient. We overview four potential data-generation processes $\{\Psi_{1,k}\}_{k=1}^4$ that are fully specified in Appendix E.2. The base-case process $\Psi_{1,1}$ is such that the seizure counts from the same patient exhibit a latent dependence structure that is rendered independent after conditioning on the covariates in the Poisson mean model. In $\Psi_{1,2}$ and $\Psi_{1,3}$, $\Psi_{1,1}$ is modified so that the latent conditional dependence structure stems from a Gaussian copula with, respectively, exchangeable and AR(1) correlation. The process $\Psi_{1, 4}$ gives rise to unstructured latent conditional dependence characterized by stronger correlations among post-baseline counts than with the pre-baseline count. Dropout is not incorporated into any of the data-generation processes.

\begin{figure}[!tb]
		\includegraphics[width = \textwidth]{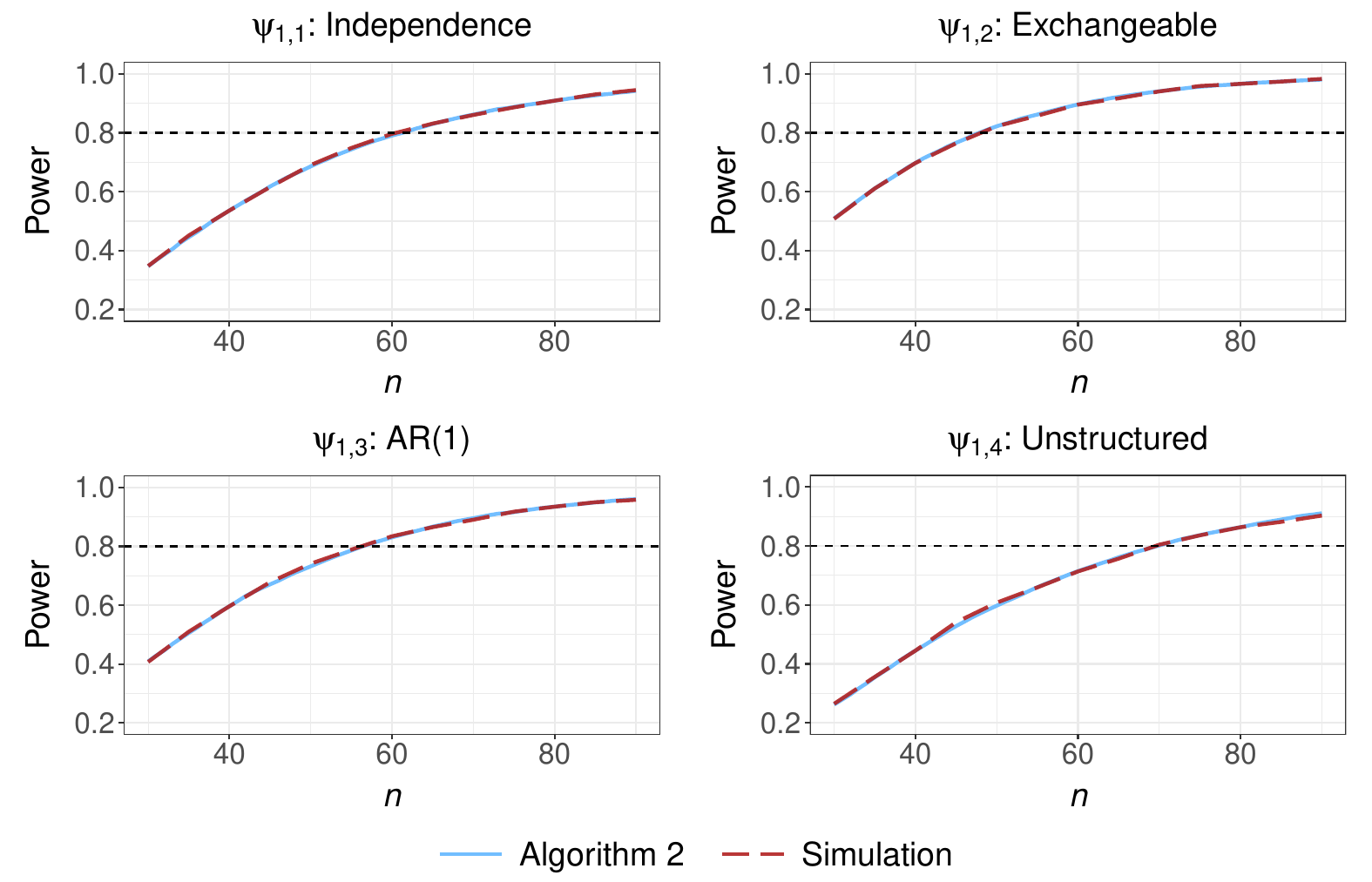} 

		\caption{\label{fig:gee} Power curves obtained using two estimation methods for the example with generalized estimating equations. The horizontal dotted line represents $1 - \beta = 0.8$. } 
	\end{figure}

    In all simulations, we estimate the standard error of the estimator for $\theta$ using sandwich estimation based on empirical residuals arising from an independence working correlation matrix. We first estimate the power curves for each $\Psi_{1,k}$ using Algorithm \ref{alg2} with $n_0 = 40$ and $n_1 = 80$ to explore sample sizes similar to that from \citet{thall1990some}. We then approximate the power curve for each $\Psi_{1,k}$ by naively simulating the sampling distribution of $p$-values for $n \in \{30, 35, \dots, 90\}$. Figure \ref{fig:gee} visualizes the power curves. 

     Although our methodology is driven by asymptotic theory, the performance of Algorithm \ref{alg2} is suitable with small to moderate sample sizes as illustrated in Figure \ref{fig:gee}. Nevertheless, we do not recommend using Algorithm \ref{alg2} to explore sample sizes with less than about $n = 20$ to $30$ ISUs. Roughly 15 seconds on a high-computing server with 72 cores were required to approximate the power curve via Algorithm \ref{alg2} for each $\Psi_{1,k}$. It took about 2 minutes to approximate the power curves in Figure \ref{fig:gee} with naive simulation. The robust sample size recommendation obtained via Algorithm \ref{alg2} is $n = 70$ since $\{n_{2,k}\}_{k=1}^4 = \{62, 48, 57, 70 \}$. We confirm via simulation that the desired type I error rate and power are reasonably attained for the recommended design in Appendix E.3. In that appendix, we also describe how the linear approximations from Algorithm \ref{alg2} could be repurposed to obtain a sample size recommendation that is robust to a weighted combination of the data-generating processes $\{\Psi_{1,k}\}_{k=1}^4$.

 \section{Discussion}\label{sec:disc}

 In this paper, we developed an efficient framework to determine sample sizes for complex hypothesis tests. The resulting sample size recommendations robustly attain the desired study power under a collection of data-generating processes that may reflect various assumptions about treatment assignment, heterogeneity, dependence, and dropout. The efficiency of this framework stems from considering a proxy for the sampling distribution of $p$-values based on large-sample theory to justify estimating the true sampling distributions at only two sample sizes per data-generating mechanism considered. This approach substantially reduces the number of simulation repetitions required for robust, computational design. Our design framework is broadly applicable to hypothesis tests facilitated via frequentist M-estimation. We accommodate one-sided and two-sided hypotheses as well as equivalence tests, in settings with clustered and non-clustered data.  

 Despite the broad scope of our design methods, this framework could be extended in several aspects.  For example, future research could develop economical and robust design methods that account for sequential analyses allowing for early termination or the multiple comparisons problem more generally. To achieve this objective, the theoretical results from this paper could be integrated with findings from work on multivariate proxy sampling distributions for the efficient but non-robust design of fully parametric Bayesian studies \citep{hagar2025fdr, hagar2025sequential}. Moreover, our design framework for robust SSD only accommodates frequentist analyses, but general Bayesian inference flexibly extends M-estimation to the Bayesian paradigm \citep{bissiri2016general}. It would be worthwhile to extend our robust design framework to accommodate such semiparametric Bayesian analyses. 

 Lastly, we acknowledge that our design methodology requires users to conduct numerical simulations for every data-generating process in $\{\Psi_{1,k}\}_{k=1}^K$. While our methodology allows users to efficiently explore the sample size space for each model $\Psi_{1,k}$, there are still computational limitations when a large number of data-generating mechanisms must be considered for robust SSD. To address these limitations, it may be possible to combine our theory related to linear approximations with surrogate modeling approaches \citep{golchi2022estimating} to efficiently consider the sample size space and the $\Psi_1$-space.

 \section*{Supplementary Material}
These materials include the proof of Theorem \ref{thm1} and additional content for the examples in Section \ref{sec:studies}. The code to conduct the numerical studies in the paper is available online: \url{https://github.com/lmhagar/RobustSSD}.


	\section*{Funding}
 
This work was supported by The University of Queensland's Health Research Accelerator initiative. 
	


\bibliographystyle{chicago}


\begin{thebibliography}{}

\bibitem[\protect\citeauthoryear{Bang and Robins}{Bang and Robins}{2005}]{bang2005doubly}
Bang, H. and J.~M. Robins (2005).
\newblock Doubly robust estimation in missing data and causal inference models.
\newblock {\em Biometrics\/}~{\em 61\/}(4), 962--973.

\bibitem[\protect\citeauthoryear{Bissiri, Holmes, and Walker}{Bissiri et~al.}{2016}]{bissiri2016general}
Bissiri, P.~G., C.~C. Holmes, and S.~G. Walker (2016).
\newblock A general framework for updating belief distributions.
\newblock {\em Journal of the Royal Statistical Society Series B: Statistical Methodology\/}~{\em 78\/}(5), 1103--1130.

\bibitem[\protect\citeauthoryear{Boyles}{Boyles}{1997}]{boyles1997estimating}
Boyles, R.~A. (1997).
\newblock Estimating common-cause sigma in the presence of special causes.
\newblock {\em Journal of Quality Technology\/}~{\em 29\/}(4), 381--395.

\bibitem[\protect\citeauthoryear{Breslow and Clayton}{Breslow and Clayton}{1993}]{breslow1993approximate}
Breslow, N.~E. and D.~G. Clayton (1993).
\newblock Approximate inference in generalized linear mixed models.
\newblock {\em Journal of the American Statistical Association\/}~{\em 88\/}(421), 9--25.

\bibitem[\protect\citeauthoryear{Chatterjee and Bose}{Chatterjee and Bose}{2005}]{chatterjee2005generalized}
Chatterjee, S. and A.~Bose (2005).
\newblock {Generalized bootstrap for estimating equations}.
\newblock {\em The Annals of Statistics\/}~{\em 33\/}(1), 414 -- 436.

\bibitem[\protect\citeauthoryear{Cohen}{Cohen}{1962}]{cohen1962statistical}
Cohen, J. (1962).
\newblock The statistical power of abnormal-social psychological research: a review.
\newblock {\em The Journal of Abnormal and Social Psychology\/}~{\em 65\/}(3), 145.

\bibitem[\protect\citeauthoryear{Davis, Martin, Stockler, Begbie, Chi, Chowdhury, Coskinas, Frydenberg, Hague, Horvath, et~al.}{Davis et~al.}{2019}]{davis2019enzalutamide}
Davis, I.~D., A.~J. Martin, M.~R. Stockler, S.~Begbie, K.~N. Chi, S.~Chowdhury, X.~Coskinas, M.~Frydenberg, W.~E. Hague, L.~G. Horvath, et~al. (2019).
\newblock Enzalutamide with standard first-line therapy in metastatic prostate cancer.
\newblock {\em New England Journal of Medicine\/}~{\em 381\/}(2), 121--131.

\bibitem[\protect\citeauthoryear{Efron}{Efron}{1982}]{efron1982jackknife}
Efron, B. (1982).
\newblock {\em The Jackknife, the Bootstrap and Other Resampling Plans}.
\newblock SIAM.

\bibitem[\protect\citeauthoryear{Faridani}{Faridani}{2025}]{faridani2025testing}
Faridani, S. (2025).
\newblock Testing for underpowered literatures.
\newblock {\em arXiv preprint arXiv:2406.13122\/}.

\bibitem[\protect\citeauthoryear{Fayers, Aaronson, Bjordal, Gr{\o}nvold, Curran, and Bottomley}{Fayers et~al.}{2001}]{fayers2001eortc}
Fayers, P., N.~K. Aaronson, K.~Bjordal, M.~Gr{\o}nvold, D.~Curran, and A.~Bottomley (2001).
\newblock {\em EORTC QLQ-C30 scoring manual}.
\newblock European Organisation for research and treatment of cancer.

\bibitem[\protect\citeauthoryear{FDA}{FDA}{2019}]{fda2019adaptive}
FDA (2019).
\newblock Adaptive designs for clinical trials of drugs and biologics — {G}uidance for industry.
\newblock Center for {D}rug {E}valuation and {R}esearch, U.S. Food and Drug Administration, Rockville, MD.

\bibitem[\protect\citeauthoryear{Fitzmaurice, Laird, and Ware}{Fitzmaurice et~al.}{2012}]{fitzmaurice2012applied}
Fitzmaurice, G.~M., N.~M. Laird, and J.~H. Ware (2012).
\newblock {\em Applied Longitudinal Analysis}.
\newblock John Wiley \& Sons.

\bibitem[\protect\citeauthoryear{Godambe}{Godambe}{1960}]{godambe1960optimum}
Godambe, V.~P. (1960).
\newblock An optimum property of regular maximum likelihood estimation.
\newblock {\em The Annals of Mathematical Statistics\/}~{\em 31\/}(4), 1208--1211.

\bibitem[\protect\citeauthoryear{Golchi}{Golchi}{2022}]{golchi2022estimating}
Golchi, S. (2022).
\newblock Estimating design operating characteristics in {B}ayesian adaptive clinical trials.
\newblock {\em Canadian Journal of Statistics\/}~{\em 50\/}(2), 417--436.

\bibitem[\protect\citeauthoryear{Green and MacLeod}{Green and MacLeod}{2016}]{green2016simr}
Green, P. and C.~J. MacLeod (2016).
\newblock {SIMR}: An {R} package for power analysis of generalized linear mixed models by simulation.
\newblock {\em Methods in Ecology and Evolution\/}~{\em 7\/}(4), 493--498.

\bibitem[\protect\citeauthoryear{Hagar, Golchi, and Klein}{Hagar et~al.}{2025}]{hagar2025sequential}
Hagar, L., S.~Golchi, and M.~B. Klein (2025).
\newblock Group sequential design with posterior and posterior predictive probabilities.
\newblock {\em arXiv preprint arXiv:2504.00856\/}.

\bibitem[\protect\citeauthoryear{Hagar and Stevens}{Hagar and Stevens}{2025a}]{hagar2025fdr}
Hagar, L. and N.~T. Stevens (2025a).
\newblock Design of {B}ayesian {A}/{B} tests controlling false discovery rates and power.
\newblock {\em arXiv preprint arXiv:2312.10814\/}.

\bibitem[\protect\citeauthoryear{Hagar and Stevens}{Hagar and Stevens}{2025b}]{hagar2024scalable}
Hagar, L. and N.~T. Stevens (2025b).
\newblock An economical approach to design posterior analyses.
\newblock {\em Journal of the American Statistical Association\/}, doi.org/10.1080/01621459.2025.2476221.

\bibitem[\protect\citeauthoryear{Huber}{Huber}{1964}]{huber1964robust}
Huber, P.~J. (1964).
\newblock Robust estimation of a location parameter.
\newblock {\em Annals of Mathematical Statistics\/}~{\em 35}, 73--101.

\bibitem[\protect\citeauthoryear{Huber}{Huber}{1973}]{huber1973robust}
Huber, P.~J. (1973).
\newblock Robust regression: asymptotics, conjectures and monte carlo.
\newblock {\em The Annals of Statistics\/}, 799--821.

\bibitem[\protect\citeauthoryear{Imai and Ratkovic}{Imai and Ratkovic}{2014}]{imai2014covariate}
Imai, K. and M.~Ratkovic (2014).
\newblock Covariate balancing propensity score.
\newblock {\em Journal of the Royal Statistical Society Series B: Statistical Methodology\/}~{\em 76\/}(1), 243--263.

\bibitem[\protect\citeauthoryear{Ito and Sugasawa}{Ito and Sugasawa}{2023}]{ito2023grouped}
Ito, T. and S.~Sugasawa (2023).
\newblock Grouped generalized estimating equations for longitudinal data analysis.
\newblock {\em Biometrics\/}~{\em 79\/}(3), 1868--1879.

\bibitem[\protect\citeauthoryear{Laird and Ware}{Laird and Ware}{1982}]{laird1982random}
Laird, N.~M. and J.~H. Ware (1982).
\newblock Random-effects models for longitudinal data.
\newblock {\em Biometrics\/}, 963--974.

\bibitem[\protect\citeauthoryear{Lehmann and Romano}{Lehmann and Romano}{2005}]{lehmann2005testing}
Lehmann, E.~L. and J.~P. Romano (2005).
\newblock {\em Testing Statistical Hypotheses}.
\newblock Springer.

\bibitem[\protect\citeauthoryear{Liang and Zeger}{Liang and Zeger}{1986}]{liang1986longitudinal}
Liang, K.-Y. and S.~L. Zeger (1986).
\newblock Longitudinal data analysis using generalized linear models.
\newblock {\em Biometrika\/}~{\em 73\/}(1), 13--22.

\bibitem[\protect\citeauthoryear{Little and Rubin}{Little and Rubin}{2019}]{little2019statistical}
Little, R.~J. and D.~B. Rubin (2019).
\newblock {\em Statistical Analysis with Missing Data}.
\newblock John Wiley \& Sons.

\bibitem[\protect\citeauthoryear{Martin, Soon, Sjoquist, Pavlakis, Goldstein, Shitara, and Simes}{Martin et~al.}{2025}]{martin2025health}
Martin, A.~J., Y.~Y. Soon, K.~M. Sjoquist, N.~Pavlakis, D.~Goldstein, K.~Shitara, and J.~R. Simes (2025).
\newblock Health-related quality-of-life outcomes with regorafenib in advanced gastric and esophagogastric junction cancer: results from the integrate trials.
\newblock {\em Gastric Cancer\/}, 1--8.

\bibitem[\protect\citeauthoryear{Pan}{Pan}{2001}]{pan2001akaike}
Pan, W. (2001).
\newblock Akaike's information criterion in generalized estimating equations.
\newblock {\em Biometrics\/}~{\em 57\/}(1), 120--125.

\bibitem[\protect\citeauthoryear{Parzen, Wei, and Ying}{Parzen et~al.}{1994}]{parzen1994resampling}
Parzen, M.~I., L.-J. Wei, and Z.~Ying (1994).
\newblock A resampling method based on pivotal estimating functions.
\newblock {\em Biometrika\/}~{\em 81\/}(2), 341--350.

\bibitem[\protect\citeauthoryear{Robins, Rotnitzky, and Zhao}{Robins et~al.}{1994}]{robins1994estimation}
Robins, J.~M., A.~Rotnitzky, and L.~P. Zhao (1994).
\newblock Estimation of regression coefficients when some regressors are not always observed.
\newblock {\em Journal of the American Statistical Association\/}~{\em 89\/}(427), 846--866.

\bibitem[\protect\citeauthoryear{Rosenbaum and Rubin}{Rosenbaum and Rubin}{1983}]{rosenbaum1983central}
Rosenbaum, P.~R. and D.~B. Rubin (1983).
\newblock The central role of the propensity score in observational studies for causal effects.
\newblock {\em Biometrika\/}~{\em 70\/}(1), 41--55.

\bibitem[\protect\citeauthoryear{Rue, Martino, and Chopin}{Rue et~al.}{2009}]{rue2009approximate}
Rue, H., S.~Martino, and N.~Chopin (2009).
\newblock Approximate {B}ayesian inference for latent {G}aussian models by using integrated nested {L}aplace approximations.
\newblock {\em Journal of the Royal Statistical Society Series B: Statistical Methodology\/}~{\em 71\/}(2), 319--392.

\bibitem[\protect\citeauthoryear{Schuirmann}{Schuirmann}{1987}]{schuirmann1987comparison}
Schuirmann, D.~J. (1987).
\newblock A comparison of the two one-sided tests procedure and the power approach for assessing the equivalence of average bioavailability.
\newblock {\em Journal of pharmacokinetics and biopharmaceutics\/}~{\em 15}, 657--680.

\bibitem[\protect\citeauthoryear{Soon, Marschner, Schou, Sweeney, Davis, Stockler, and Martin}{Soon et~al.}{2025}]{soon2025target}
Soon, Y.~Y., I.~C. Marschner, I.~M. Schou, C.~J. Sweeney, I.~D. Davis, M.~R. Stockler, and A.~J. Martin (2025).
\newblock Target trial emulation of early docetaxel and enzalutamide for metastatic hormone-sensitive prostate cancer.
\newblock {\em BJU International\/}, doi:10.1111/bju.70065.

\bibitem[\protect\citeauthoryear{Stroup, Ptukhina, and Garai}{Stroup et~al.}{2024}]{stroup2024generalized}
Stroup, W.~W., M.~Ptukhina, and J.~Garai (2024).
\newblock {\em Generalized Linear Mixed models: Modern Concepts, Methods and Applications}.
\newblock Chapman and Hall/CRC.

\bibitem[\protect\citeauthoryear{Sweeney, Martin, Stockler, Begbie, Cheung, Chi, Chowdhury, Frydenberg, Horvath, Joshua, et~al.}{Sweeney et~al.}{2023}]{sweeney2023testosterone}
Sweeney, C.~J., A.~J. Martin, M.~R. Stockler, S.~Begbie, L.~Cheung, K.~N. Chi, S.~Chowdhury, M.~Frydenberg, L.~G. Horvath, A.~M. Joshua, et~al. (2023).
\newblock Testosterone suppression plus enzalutamide versus testosterone suppression plus standard antiandrogen therapy for metastatic hormone-sensitive prostate cancer (enzamet): an international, open-label, randomised, phase 3 trial.
\newblock {\em Lancet Oncol\/}~{\em 24\/}(4), 323--334.

\bibitem[\protect\citeauthoryear{Thall and Vail}{Thall and Vail}{1990}]{thall1990some}
Thall, P.~F. and S.~C. Vail (1990).
\newblock Some covariance models for longitudinal count data with overdispersion.
\newblock {\em Biometrics\/}, 657--671.

\bibitem[\protect\citeauthoryear{van~der Laan and Robins}{van~der Laan and Robins}{2003}]{laan2003unified}
van~der Laan, M.~J. and J.~M. Robins (2003).
\newblock {\em Unified Methods for Censored Longitudinal Data and Causality}.
\newblock Springer.

\bibitem[\protect\citeauthoryear{van~der Vaart}{van~der Vaart}{1998}]{vaart1998bvm}
van~der Vaart, A.~W. (1998).
\newblock {\em Asymptotic Statistics}.
\newblock Cambridge Series in Statistical and Probabilistic Mathematics. Cambridge University Press.

\bibitem[\protect\citeauthoryear{Wang and Chen}{Wang and Chen}{2001}]{wang2001augmented}
Wang, C. and H.~Y. Chen (2001).
\newblock Augmented inverse probability weighted estimator for {C}ox missing covariate regression.
\newblock {\em Biometrics\/}~{\em 57\/}(2), 414--419.

\bibitem[\protect\citeauthoryear{Wellek}{Wellek}{2010}]{wellek2010testing}
Wellek, S. (2010).
\newblock {\em Testing Statistical Hypotheses of Equivalence and Noninferiority}.
\newblock Chapman and Hall/CRC.

\bibitem[\protect\citeauthoryear{White}{White}{1980}]{white1980heteroskedasticity}
White, H. (1980).
\newblock A heteroskedasticity-consistent covariance matrix estimator and a direct test for heteroskedasticity.
\newblock {\em Econometrica: Journal of the Econometric Society\/}, 817--838.

\bibitem[\protect\citeauthoryear{Zeger and Liang}{Zeger and Liang}{1986}]{zeger1986longitudinal}
Zeger, S.~L. and K.-Y. Liang (1986).
\newblock Longitudinal data analysis for discrete and continuous outcomes.
\newblock {\em Biometrics\/}, 121--130.

\end{thebibliography}

\end{document}